# Enhancing IoT Security Against DDoS Attacks through Federated Learning

Ghazaleh Shirvani[1], Saeid Ghasemshirazi[2], Mohammad Ali Alipour[3]

[1]Carleton University, Ottawa, Ontario, Canada, saeidghshirazi@gmail.com

[2]Carleton University, Ottawa, Ontario, Canada, ghazaleh.sh3p@gmail.com

[3]Institute of Science and High Technology and Environmental Science, Kerman, Iran, m.alialipour77@gmail.com

## Abstract

The rapid proliferation of the Internet of Things (IoT) has ushered in transformative connectivity between physical devices and the digital realm. Nonetheless, the escalating threat of Distributed Denial of Service (DDoS) attacks jeopardizes the integrity and reliability of IoT networks. Conventional DDoS mitigation approaches are ill-equipped to handle the intricacies of IoT ecosystems, potentially compromising data privacy. This paper introduces an innovative strategy to bolster the security of IoT networks against DDoS attacks by harnessing the power of Federated Learning that allows multiple IoT devices or edge nodes to collaboratively build a global model while preserving data privacy and minimizing communication overhead. The research aims to investigate Federated Learning's effectiveness in detecting and mitigating DDoS attacks in IoT. Our proposed framework leverages IoT devices' collective intelligence for real-time attack detection without compromising sensitive data. This study proposes innovative deep autoencoder approaches for data dimensionality reduction, retraining, and partial selection to enhance the performance and stability of the proposed model. Additionally, two renowned aggregation algorithms, FedAvg and FedAvgM, are employed in this research. Various metrics, including true positive rate, false positive rate, and F1-score, are employed to evaluate the model. The dataset utilized in this research, N-BaIoT, exhibits non-IID data distribution, where data categories are distributed quite differently. The negative impact of these distribution disparities is managed by employing retraining and partial selection techniques, enhancing the final model's stability. Furthermore, evaluation results demonstrate that the FedAvgM aggregation algorithm outperforms FedAvg, indicating that in non-IID datasets, FedAvgM provides better stability and performance.

# 1. Introduction

The proliferation of the IoT has revolutionized numerous industries, enabling seamless connectivity and communication between physical devices and the digital world[1]. With IoT's rapid expansion and increasing adoption in critical applications like healthcare, transportation, and smart cities, ensuring the security and resilience of IoT networks becomes paramount[2]. Among the various security threats that plague IoT ecosystems, DDoS attacks pose a significant challenge, capable of disrupting IoT services and causing substantial financial and reputational losses[3]. DDoS attacks overwhelm a target network with a deluge of traffic, rendering its resources unavailable to legitimate users. Traditional DDoS mitigation approaches, predominantly based on centralized cloud-based solutions, struggle to cope with the scale and complexity of IoT networks [4]. Furthermore, these conventional methods may raise privacy concerns and exhibit limitations in addressing new attack vectors specifically tailored to exploit IoT devices and protocols [5].

In response to these challenges, this paper proposes a novel approach to secure IoT networks from DDoS attacks through the application of Federated Learning. Federated Learning is a decentralized machine learning technique that allows multiple IoT devices or edge nodes to collaboratively build a global model while preserving data privacy and minimizing communication overhead [6]. By leveraging the distributed intelligence of IoT devices, Federated Learning holds promise in addressing the security needs of large-scale and heterogeneous IoT networks [7]. The primary objective of this research is to explore the efficacy of Federated Learning as a mechanism for detecting and mitigating DDoS attacks in IoT environments. By harnessing the collective knowledge of IoT devices, our proposed framework aims to enable real-time detection and timely response to DDoS attacks without compromising sensitive data privacy[8]. Through this study, we seek to contribute to the growing body of research on securing IoT ecosystems and fostering advancements in Federated Learning applications.

This paper is structured as follows: Section 2 offers a background on the fundamental concepts related to IoT networks, DDoS attacks, and Federated Learning. In Section 3, we provide a comprehensive review of the existing literature on IoT security and DDoS mitigation techniques. Section 4 discusses the motivation and aims behind this study. Building on this foundation, Section 5 outlines our proposed framework for securing IoT networks using Federated Learning. Section 6 presents the experimental setup and results to demonstrate the effectiveness of our approach. We then analyze and discuss the findings. Conclusion and future research directions are placed in Sections 7 and 8.

By combining the power of IoT with Federated Learning, we aim to pave the way for a resilient and secure IoT ecosystem capable of withstanding sophisticated DDoS attacks and safeguarding critical services for a wide range of applications. As IoT continues to grow and permeate into various domains, addressing its security challenges becomes an imperative step toward realizing the full potential of this transformative technology.

## 2. Background

### 2.1.Internet of Things and Security Challenges

The advent of the Internet of Things has ushered in a new era of interconnected devices, revolutionizing various domains, including healthcare, transportation, and smart cities [9]. The seamless integration of physical objects with digital networks offers unprecedented opportunities for data-driven decision-making and automation [10]. However, this rapid proliferation of IoT technologies has also brought forth significant security challenges. Among the various threats to IoT ecosystems, Distributed Denial of Service attacks have emerged as a prominent concern [11].

### 2.2.Distributed Denial of Service Attacks on IoT Networks

DDoS attacks target network infrastructures by overwhelming them with a barrage of malicious. traffic, rendering the services inaccessible to legitimate users. In traditional settings, DDoS attacks have been predominantly addressed using centralized cloud-based mitigation approaches [12]. Nonetheless, the distinctive characteristics of IoT networks, such as heterogeneity, resource constraints, and the distributed nature of devices, pose formidable challenges to the efficacy of conventional DDoS mitigation strategies [13].

### 2.3.Federated Learning as a Potential Solution

Federated Learning (FL) is a decentralized machine learning technique that has gained prominence in recent years. It allows collaborative model training without centralizing raw data [14]. FL allows IoT devices or edge nodes to participate collaboratively in building a global model while preserving the privacy of individual data sources. This distributed intelligence paradigm aligns well with the inherent characteristics of IoT networks, potentially offering a novel solution to mitigate DDoS attacks [15].

### 2.4.Deep Autoencoder

A deep autoencoder is a type of deep neural network model used for data transformation and compression. It is employed to learn low-dimensional data representations, aiming to capture essential information from the original data [16]. The primary objective of this model is to convert input data into a low-dimensional latent space where crucial and fundamental data information is encapsulated. This low-dimensional latent space can serve as a meaningful and informative representation of data for various tasks such as data dimensionality reduction, noise reduction, learning important features, and extracting conceptual information from the data [17].

In a deep autoencoder, data is transformed into a latent space through layers of hidden nodes in the network. Subsequently, the data is reconstructed from this latent space back to its original form while preserving significant features. The neural network of a deep autoencoder consists of hidden layers (Figure 1) that strive to create a low-dimensional representation of the data, transferring the most important data information while retaining their meaningful characteristics[18].

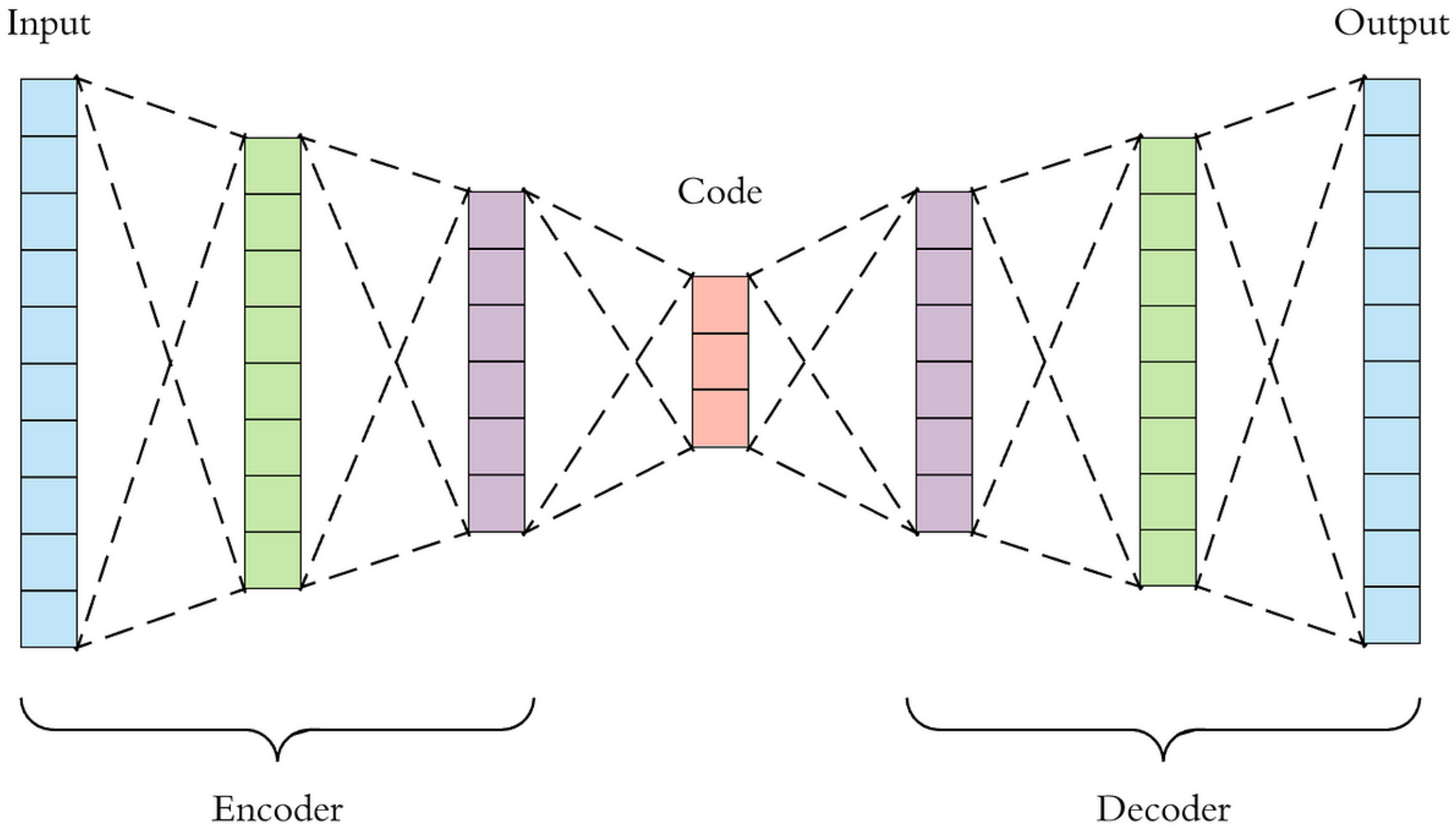

**Figure 1.** Deep Autoencoder Architecture

One of the key advantages of deep autoencoders is their ability to autonomously learn a suitable low-dimensional representation of data without requiring precise labels for the data. As a result, these models can be employed in an unsupervised manner for tasks such as data dimensionality reduction, noise elimination, pattern identification, and anomaly detection in sensor data and various devices. Through this model, it becomes possible to extract conceptual information from IoT data and automatically detect normal patterns and behaviors without prior knowledge, effectively providing the capability to identify potential attacks and anomalies in IoT networks and systems [19].

## 3. Related work

In the realm of IoT security and DDoS attack mitigation, several research studies have focused on exploring different approaches to enhance the resilience of IoT networks. This section presents a comprehensive review of related works and highlights existing methods' key contributions and limitations.

### 3.1.DDoS Mitigation Techniques for IoT

Various traditional DDoS mitigation techniques have been adapted for IoT environments to protect against volumetric and application-layer attacks [20]. Researchers have proposed rate-limiting mechanisms, traffic filtering, and blacklisting to mitigate the impact of DDoS attacks on IoT devices. While these methods are effective in specific scenarios, they may not be well-suited for IoT networks' dynamic and resource-constrained nature, potentially leading to high false positive rates and inadequate protection against sophisticated attacks [21], [22].

### 3.2.Machine Learning-Based DDoS Detection

Machine learning techniques have shown promise in detecting DDoS attacks in conventional networks. Scientists have investigated supervised learning algorithms, including Support Vector Machines (SVM), Random Forests (RF), and Neural Networks, with the purpose of detecting malicious traffic patterns. While these methods have demonstrated reasonable accuracy in detecting DDoS attacks, they often require centralized data aggregation, posing privacy and scalability concerns in IoT environments [23].

### 3.3.Federated Learning for IoT Security

Federated learning, which maintains training data locally while multiple devices train a global learning model collectively, is a category of distributed machine learning algorithms [24]. In particular, IoT devices ensure data privacy by training their local machine learning models with local data and sharing the trained local model parameters rather than the raw data.

In the realm of IoT communications, Federated Learning assumes vital roles. It conserves energy and wireless resources by exchanging limited model parameters instead of transmitting extensive training data [25]. The orchestration of local model training concurrently curtails transmission latency, amplifying the efficiency of data flow. Crucially, data privacy is maintained since data resides on devices, and only local parameters are shared [26]. Diverse training methods for multiple classifiers on distributed datasets enhance accuracy, particularly in large-scale domains. Moreover, federated learning's inherent scalability accommodates growing data volumes, providing a solution to challenges like algorithm complexity and memory limitations. The federated learning paradigm involves an iterative process for refining the global model. The process includes three stages: initial model provisioning, local model training, and aggregation of local models. This approach, depicted in Figure 2, showcases a continuous learning process, repeating earlier stages to maintain the global model for all participants in the current cycle.

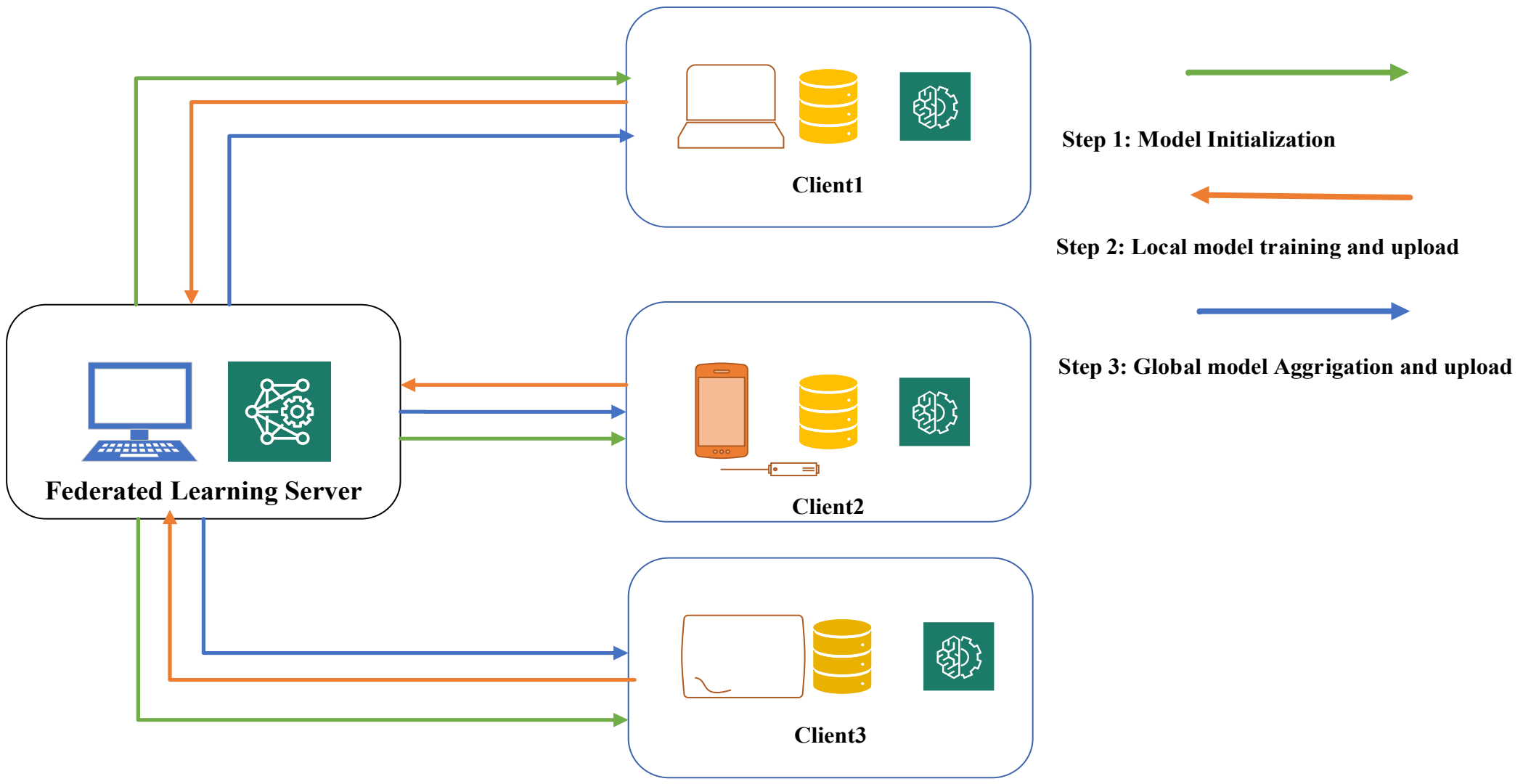

**Figure 2.** Representation of decentralized federated learning

Federated Learning has become known as a decentralized framework in the field of machine learning, seamlessly integrating with the complex infrastructure of IoT networks. Noteworthy investigations, as presented in Table 1, have demonstrated the application of Federated Learning to fortify IoT security, encompassing domains such as intrusion detection and anomaly detection [27]. While Federated Learning addresses privacy and communication overhead, its application in mitigating DDoS attacks in IoT contexts requires further evaluation and optimization. This area remains relatively new and warrants in-depth exploration.

**Table 1.** Comparison of the related articles to detection of DDoS attacks using federated learning

| **Dataset** | **Implementation** | **Machine Learning** | **Aggregation Method** | **Reference** |
|---|---|---|---|---|
| Simulated traffic | Not Available | GRU | FedAvg | [28] |
| KDDCup99 | Not Available | MLP, DT, SVM, RF | FedAvg | [29] |
| Generated | Not Available | GRU | FedAvg | [30] |
| NSL-KDD | TensorFlow | MLPs | FedAvg | [31] |
| CSE-CIC-IDS2018 | TensorFlow | NN | FedAvg | [32] |
| CICIDS2017, ISCX Botnet 2014 | TensorFlow | Binarized NN | signSGD | [33] |
| KDD, NSL-KDD, | Not Available | Deep belief network | FedAvg | [34] |
| BoT-IoT | Not Available | NN | FedAvg | [35] |
| CIC-ToN-IoT | IBMFL | Logistic regression | FedAvg, Fed+ | [36] |
| Modbus dataset | Pytorch/PySyft | GRU | FedAvg | [37] |

### 3.4.Hybrid Approaches

To tackle the challenges of DDoS attacks in IoT, some researchers have proposed hybrid approaches that combine traditional mitigation techniques with machine learning-based detection methods [38]. These hybrid approaches aim to leverage the ability of both approaches to achieve better adaptability and higher accuracy to the dynamic IoT environment. Although the hybrid models have demonstrated encouraging outcomes, they frequently necessitate substantial computational resources and may not be appropriate for IoT devices with severe resource limitations [39].

## 4. Motivations and Aims

Recently, the advancement of IoT has brought significant attention to the issues related to network traffic attacks. However, due to the inherent simplicity of these devices, they are easily exploited for denial-of-service attacks. IoT devices can be utilized as bots to launch DDoS attacks. The rapid growth of IoT devices, often with lower capabilities compared to desktop computers, has led to a surge in IoT botnet attacks. Botnet attacks refer to a type of DDoS attack where an attacker leverages a large number of compromised IoT devices (referred to as attacking bots) to target a specific victim rapidly. Detecting such attacks is challenging as the compromised devices continue to function normally, and device owners may remain unaware of their devices' involvement in an attack [40].

As outlined in the related work, for enhancing network security against such attacks and the methods for identifying them, most existing approaches involve the collection of data, which is then centralized for training machine learning models [41]. The primary drawbacks of such approaches include data leakage risks and the inability to detect attacks promptly. This research utilizes federated learning to address these issues and increase the security and universality of intrusion detection models for DDoS attacks in IoT networks [42].

In the course of this research, the challenges associated with data leakage, handling non-independent and heterogeneous IoT data distributions, model training efficiency, and performance will be thoroughly examined. The primary reasons for choosing the proposed federated learning model for this subject are as follows:

1. Given the diverse characteristics of each user and the potential impact of different types of devices on the overall model performance, federated learning becomes highly relevant, promising considerable improvement in overall model efficacy.
2. Federated learning has the capacity to preserve user privacy, as participants are only required to share the trained model's parameters, eliminating the need to transmit or share their actual sensitive data.

In summary, the surge in IoT-based DDoS attacks, coupled with the unique characteristics and distribution of IoT data necessitates innovative approaches to intrusion detection. This paper aims to address these challenges through the application of federated learning, with the objectives of enhancing model performance, preserving user privacy, and providing a comprehensive solution to combat the emerging threat landscape in IoT networks.

## 5. Proposed Methodology

To achieve the research objective, we implemented a deep autoencoder model in conjunction with federated learning for the purpose of attack detection. Federated learning can be an effective method for training deep learning models to analyze data across a diverse range of devices while safeguarding the privacy of the device users, according to research from multiple disciplines.

Figure 4 presents an in-depth depiction of the architectural configuration of the model that is currently in consideration. In a nutshell, the data preprocessing for each device that is taking part in the training process is done in the exact same way that it was done for the strategy that relies on local detection. Once the federated learning process has been initiated, the training will start happening simultaneously across all devices. During the training process, a deep autoencoder model (referred to as the "local model" throughout this document and prominently displayed in Figure 3) is utilized within each device to gain knowledge from the device's own local data. After that, each device will transmit the model weights it has calculated to the central server. Using the aggregation function (either FedAvg or FedAvgM), the global model performs processing on the weights and then makes any necessary adjustments in the central server. At the very end of the procedure, every deep autoencoder model on every device receives the freshly aggregated weights and then begins the process of training.

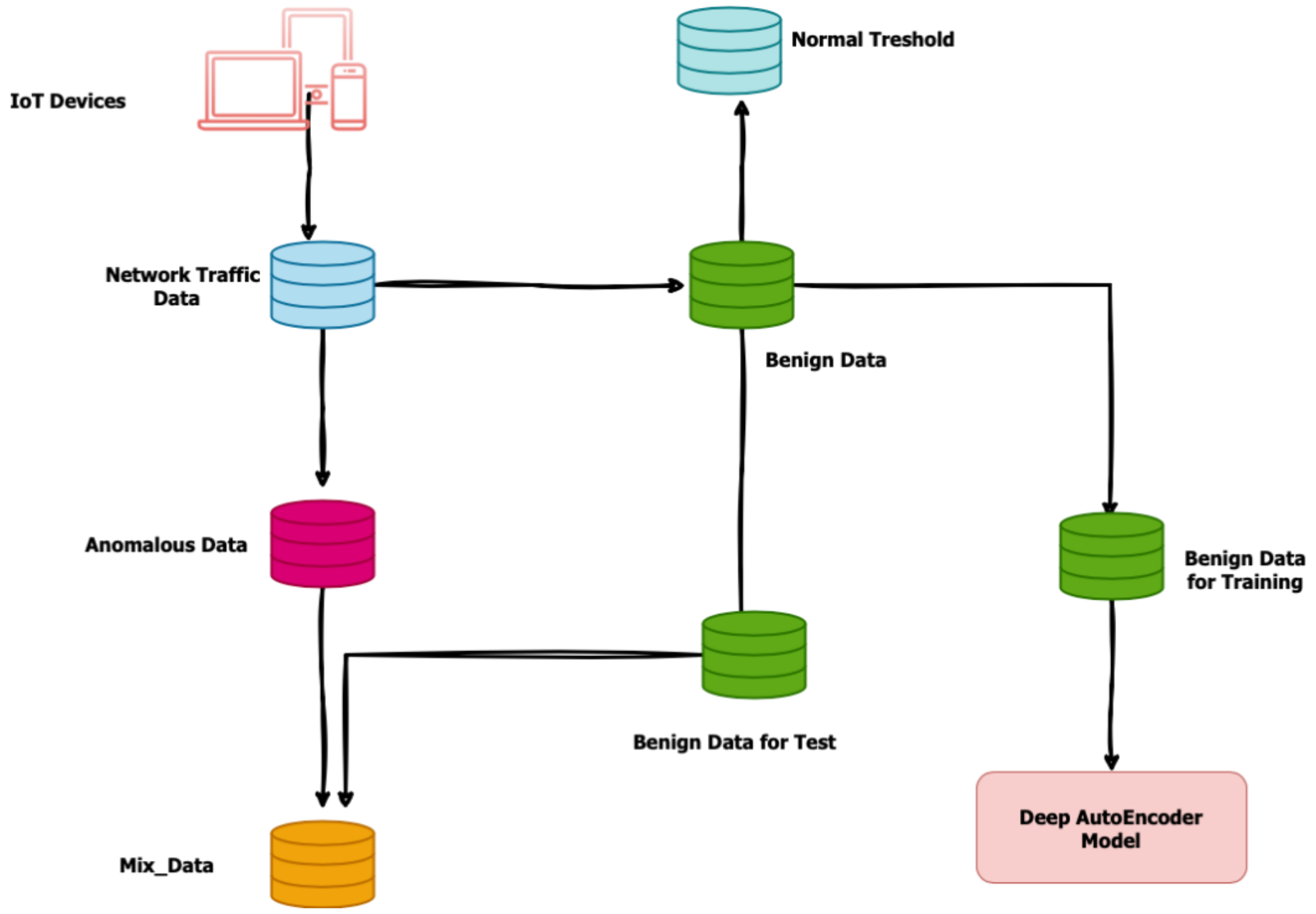

**Figure 3.** Proposed Local Model Architecture

The federated learning stage encompasses three stages: forward propagation, sending weights to the global server for updating, and backpropagation. In federated learning, these three processes are referred to as a communication round. Multiple communication rounds are employed to update the weights, with an increased number of these rounds ensuring improved performance of the federated model.

Furthermore, an additional stage, known as the retraining process, has been incorporated into the architecture of the proposed model to manage heterogeneous datasets. This ensures that the model remains efficient and practical for real-life environments. The retraining process implies that client models are additionally trained prior to the aggregation of weights in the global server. The data for this process is randomly selected from the training data associated with each device.

To improve the proposed model's stability and efficiency, a partial selection mechanism has been incorporated into the federated model architecture. This entails the arbitrary selection of a subset of client devices for practical training purposes during each communication round, as opposed to employing all devices. After the federated model has been trained, it is possible to apply the trained model directly to both the specified devices and newly acquired devices.

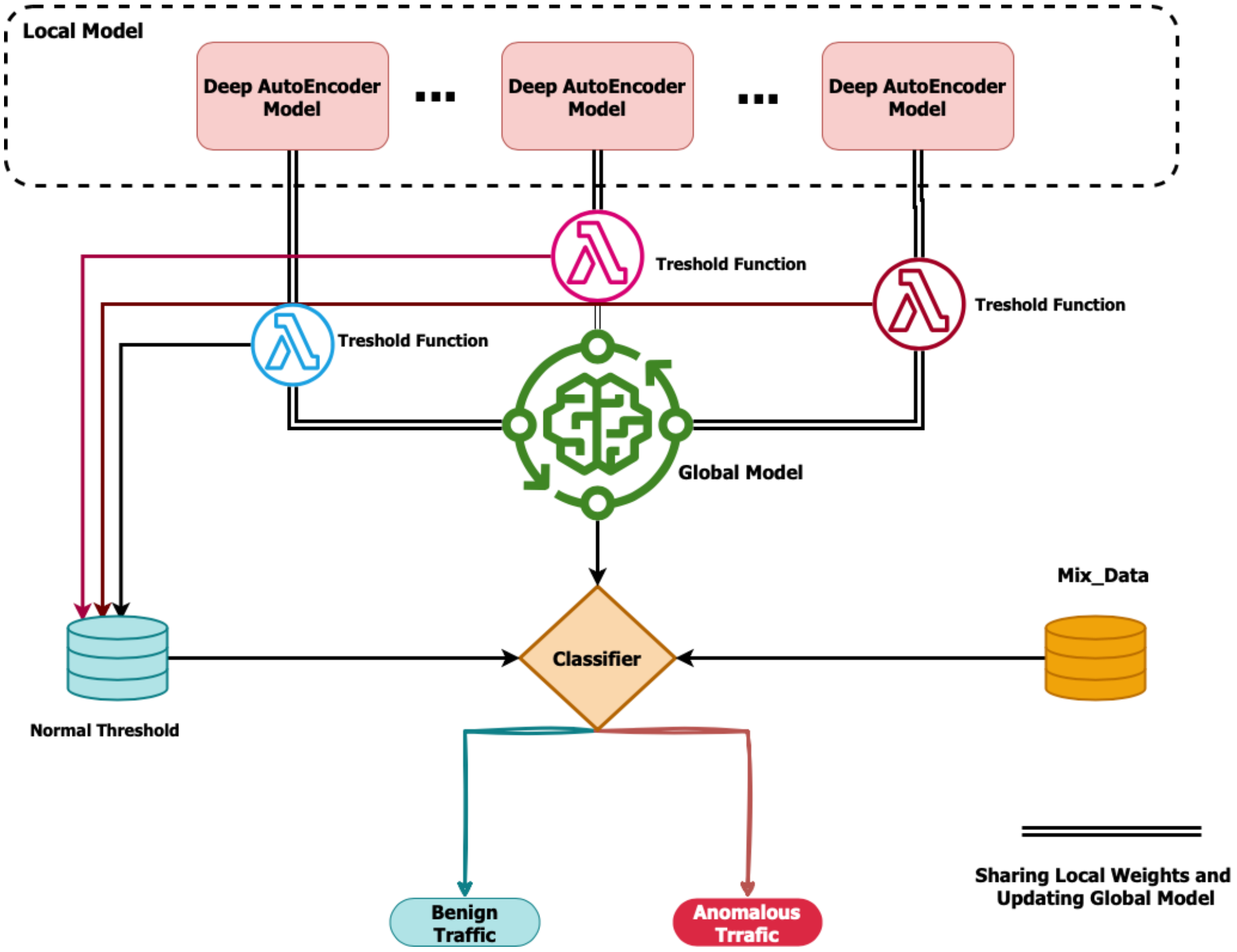

**Figure 4.** Proposed Global Model Architecture

**Retraining Process**

The Dataset employed in this study exhibits non-homogeneity, characterized by an independent and non-identical distribution (non-IID) [43]. This poses a challenge in the federated learning model, wherein settings are contingent upon the distribution of class labels and client data attributes, potentially influencing training time and overall model accuracy. In the retraining mechanism, the process of retraining the models is delineated, taking place after the defined number of communication rounds between the central server and clients. The prime objective of the process is to retrain the deep learning models through exposure to new data collected from preceding communication rounds.

In each communication round between the central server and clients, the deep learning models are trained by clients using the data at their disposal. Subsequent to the interaction with clients, the model weights are dispatched to the central server for a comprehensive retraining phase, wherein all data and model weights are utilized to facilitate retraining. The retraining process follows this. During this stage, the deep learning models undergo retraining using foundational data and a specified number of retraining epochs. Upon the culmination of the retraining process, the training error of the models in each retraining epoch is computed and stored. Eventually, the sum of retraining errors for these epochs is divided by the number of selected clients, thereby deriving an average error for each client, which is subsequently stored in a repository.

This section plays a pivotal role in enhancing and optimizing deep learning models for intrusion detection, fostering model improvement through new data, and adaptation to novel conditions within the IoT environment. Additionally, this method optimizes the model towards achieving more accurate and reliable results in threat detection.

**Partial Selection Mechanism**

In conjunction with the implementation of communication cycles to update weights and improve the overall performance of the federated learning model, the partial client selection mechanism is utilized to alleviate the influence of diverse datasets on the model. During the partial selection procedure, subsets of training devices are selected at random to be included in the training process. As opposed to selecting all nine clients for global model training in each communication round, this study selects four out of nine clients for such training.

Initially, the selected clients in each communication round update the global model's weights. In the subsequent phase, the trained models are transferred to the selected client devices, which then update and enhance the models using their localized data. Subsequently, the training error is computed and stored in this phase. After the client devices update the models, the retraining process (previously explained) commences at the central server. This process serves as an improvement endeavor, enabling the models to be updated with various and noisy data, thereby enhancing their accuracy and generalization capability. At the conclusion of each communication round, the updated models from participating devices are amalgamated with the global model. This amalgamation results in the enhancement of the global model with new insights from diverse devices, thus enabling improved detection and prediction of threats and intrusions.

The partial selection approach in federated learning, amalgamating the contributions of diverse devices and improving model learning, equips us to address the challenges of diversity and data multiplicity in IoT network environments. This approach enhances the accuracy and efficiency of models in detecting threats and intrusions and fosters an overall improvement in the federated learning system's performance.

**Model Detection Approach**

The entire process of evaluation and classification of the proposed model is encapsulated within a single function. To evaluate the accuracy and efficacy of the model in detecting threats and attacks on mixed data, the model is initially configured in the evaluation mode, ensuring its placement in the evaluation context following training.

Subsequently, employing a designated threshold, decisions are made regarding the normalcy or abnormality of each data instance. In other words, the proposed model ultimately furnishes a binary classifier for distinguishing normal data instances from abnormal ones. A variety of evaluation metrics are calculated using both actual and predicted labels. These metrics include but are not limited to accuracy, precision, recall, F1 score, true positive rate (TPR), and false positive rate (FPR). The computation of these metrics is predicated on the tally of both predicted and actual instances.

**Threshold Computation for Each Device**

Following model training and extraction of its Mean Squared Error (MSE), a threshold (tr) is employed to differentiate between normal and abnormal observations. This threshold determination is founded on the Equation (1).

$$\mathrm{tr} = \mathrm{MSE}_{\mathrm{benign_tr}} + \mathrm{std}_{(\mathrm{MSE}_{\mathrm{benigntr}})} \quad (1)$$

In (1), $\mathrm{MSE}_{\mathrm{benign_tr}}$ signifies the mean squared error of the model on training data pertaining to normal traffic or benign data, while $\mathrm{std}_{(\mathrm{MSE}_{\mathrm{benigntr}})}$ represents the standard deviation of error scores. If the observed value surpasses this threshold, the data is labeled as abnormal; otherwise, it is considered normal. In essence, this method utilizes error comparison to a specified metric, facilitating anomaly detection and aiding the model's decision-making process regarding data types.

**5.1. Utilized Hyperparameters in the Proposed Model**

Hyperparameters used for the proposed model are tuned at the values mentioned in Table 2. These hyperparameters are crucial in configuring the Federated Learning framework and fine-tuning the model's performance in detecting and mitigating DDoS attacks in IoT environments. Proper selection and tuning of these hyperparameters are essential to achieving optimal results and ensuring the robustness of the FL-based framework across diverse IoT network scenarios.

**Table 2.** Model Hyperparameters Setting

| **Hyperparameter** | **Description** | **Value** |
|---|---|---|
| num_clients | The quantity of IoT devices (clients) that are members of the FL framework. | 9 |
| num_selected | The quantity of clients chosen for every iteration of model aggregation. | 4 |
| batch_size | Batch size is used for training the local models on each client. | 128 |
| baseline_num | The number of baseline models used for comparison and validation purposes. | 1000 |
| num_rounds | The number of rounds of model aggregation during the FL training process. | 4 |
| epochs | In one round, the number of epochs is used for training the local models on each client. | 10 |
| retrain_epochs | The number of epochs used for retraining the local models after each round. | 10 |
| Optimizer | The optimization algorithm used during the FL training process (e.g., SGD, Adam). | SGD |
| Learning Rate | The initial learning rate for updating the model parameters during training. | 0.012 |
| weight_decay | The regularization term is applied to the model parameters to prevent overfitting. | e-05 |
| momentum | The momentum factor is used in optimization algorithms like SGD with momentum. | 0.9 |

# 6. Evaluation of the Proposed Method

## 6.1.Dataset

This study utilizes the N-BaIoT dataset to assess the efficacy of our proposed FL framework in identifying and mitigating DDoS attacks in IoT environments. The N-BaIoT dataset is a renowned and commonly utilized dataset in IoT security, specifically created for assessing DDoS attack detection methods. The N-BaIoT dataset comprises a diverse set of network traffic data collected from real-world IoT devices and represents various IoT applications and scenarios. It encompasses both legitimate and DDoS attack traffic, allowing us to simulate realistic IoT network environments [46]. The Dataset is characterized by its large-scale and heterogeneous nature, mirroring the complexities of real-world IoT networks.

The utilization of the proposed dataset in studies pertaining to the security of IoT carries substantial significance. The aforementioned resource functions as a significant repository of empirical data that can be utilized to analyze and assess the performance of intrusion detection models in the face of IoT botnet attacks. The availability of this information allows for the analysis and evaluation of innovative approaches within the field of IoT network security, hence supporting the enhancement of security measures for IoT devices. Each of the feature headers outlined in Table 3 presents a comprehensive discussion of their respective qualities.

The utilization of mean and standard deviation in the study of network traffic and identification of IoT botnet attacks is of paramount importance inside the dataset. These statistics provide vital information about the population of network traffic and its characteristics, which is extremely useful for detecting and predicting Internet of Things botnet attacks.

Finally, the comprehensive ratio of normal data to anomalous data in this Dataset after preprocessing corresponds to Fig. 5. As elucidated in the preceding sections, our Dataset in this study is notably imbalanced due to being collected from real IoT traffic. Given the considerably lower population of normal data instances in relation to anomalous data instances, it is clear that the challenge of accurately training a model for high-precision attack detection can be understood.

**Table 3.** Table 3. Data attribute header in N-BaIoT dataset

| Feature | Signature Format | Definition |
|---|---|---|
| Flow Aggregation | H | Summarized statistics of recent traffic from the source host (IP) of this packet. |
| | HH | Summarized statistics of recent traffic from the source host (IP) to the destination host. |
| | HpHp | Summarized statistics of recent traffic from the source host + port (IP) to the destination host + port. Example: 192.168.4.2:1242 -> 192.168.4.12:80 |
| | HH_jit | Summarized statistics of recent jitter of traffic from the source host (IP) to the destination host. |
| Time Window (Lambdajit decay | L5، L3، L1 | How much of the recent flow history is considered for these statistics |
| factor used in exponential decay) | | |
| Extracted Flow Statistics | weight | Flow weight (can be seen as the count of observed items in recent history) |
| | Radius | The sum of the squared variance of two flows |
| | Value | The sum of the squared means of two flows |
| | cov | A covariance approximately between two flows |
| | PCC | Approximate Pearson correlation coefficient between two flows |
| Mean | Mean | The average value of a dataset. To calculate the mean, all numbers in the Dataset are summed, and the result is divided by the total number of values. Mean is an important measure and indicator of data concentration. |
| Standard Deviation | Standard Deviation | Standard deviation indicates the spread of data from the mean. By calculating the standard deviation, we can see how much the data deviates from the mean and how distinct they are from each other. A larger standard deviation indicates greater data dispersion, while a smaller standard deviation indicates tighter clustering of data around the mean. |

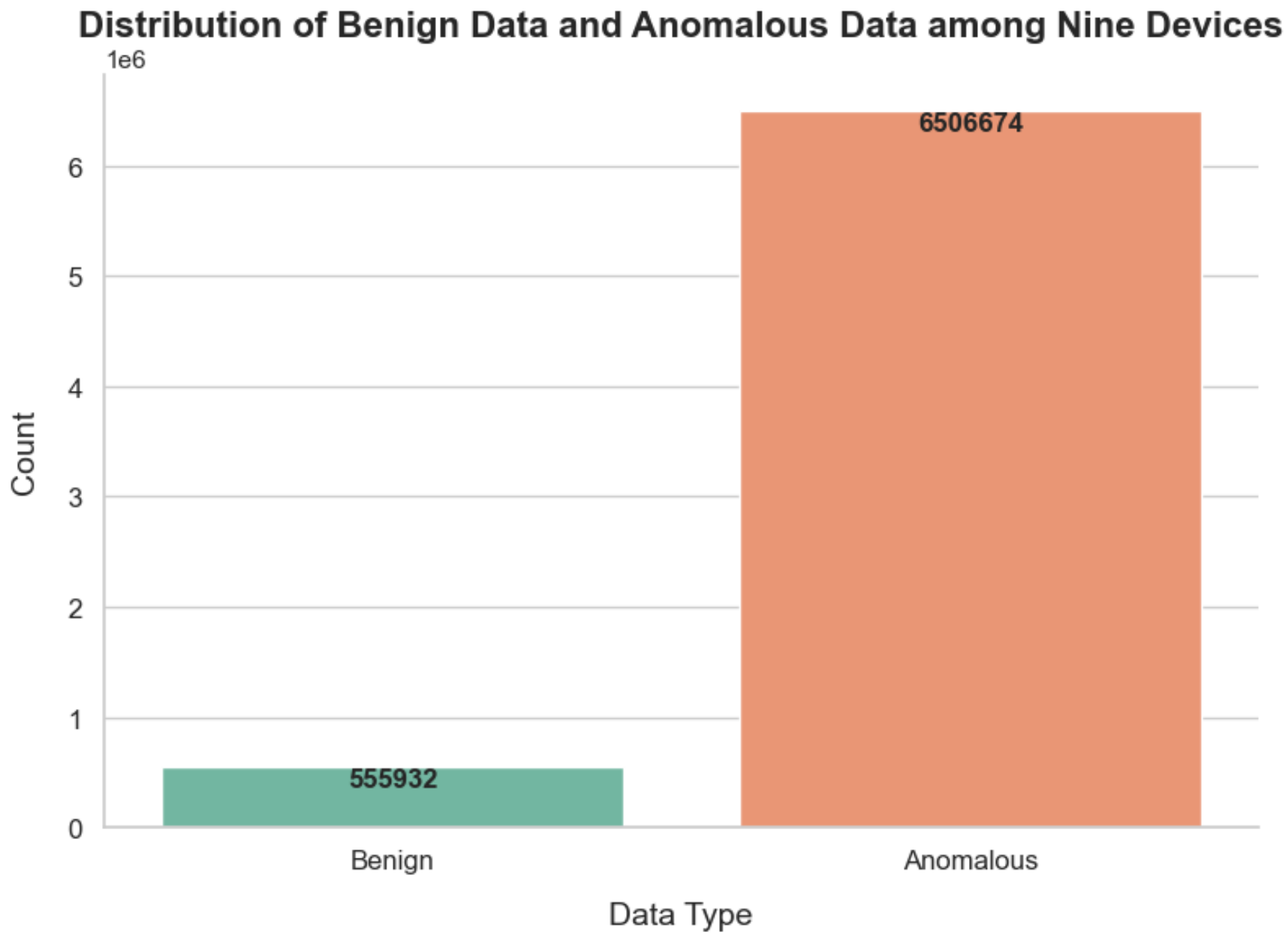

**Figure 5.** The ratio of Anomalous Data to Benign Data in N-BaIoT Dataset

## 6.2.Data Preprocessing and Cleaning

In the selected Dataset, a total of 7,062,606 data samples with 115 features are available in CSV format files. Initially, the CSV files need to be loaded, which might occupy all memory space and lead to potential errors. To mitigate potential risks and facilitate future detection processes, for each IoT device, relevant CSV files have been combined and transformed into the Parquet format. In this process, a new feature called "Type" has been introduced to signify whether the data is legitimate or compromised by Botnet attacks. Specifically, if the type is "benign," it indicates valid data. However, if the type is "scan," representing Mirai attacks, or "Combo," signifying Gafgyt attacks, it indicates that the data is compromised. Prior to consolidation, the names and order of features have been examined to ensure consistency among the Parquet files. After preprocessing, our IoT devices have a single Parquet file, each containing 116 features. The shape of each device is represented in Table 4.

Within this research phase, we diligently undertake the critical tasks of data cleansing and preprocessing on the N-BaIoT dataset. These essential steps are undertaken to ensure that the Dataset is suitably primed for the subsequent utilization of Federated Learning in the context of DDoS attack detection. The ensuing sections provide a comprehensive exposition of the meticulous procedures employed during the preprocessing and data-cleaning stages, as elegantly elucidated in Figure 6.

1) Removing Invalid Data:

Initially, incomplete or invalid data are removed from the Dataset. Some data might be incomplete or invalid due to errors in data collection or recording. Such data needs to be entirely eliminated from the Dataset to enhance the accuracy and quality of the anomaly detection model.

2) Eliminating Duplicate Data:

There could be duplicated data within the Dataset that could negatively impact the model's performance. Hence, DDoS attacks present as duplicates in the Dataset need to be removed to prevent inappropriate repetition and enable the model to achieve higher accuracy in DDoS attack detection.

3) Data Separation:

First, the first column (ID) of the data is removed. Then, the data is categorized based on the labels "benign" and "anomalous." Data labeled as "benign" are assigned the label 0, while data labeled as "anomalous" are assigned the label 1.

4) Data Splitting:

From the "benign" data, three partitions are randomly separated, each accounting for 1/3 of the total benign data instances. The first partition is for model training, the second for threshold calculation, and the third for model testing. Here, the threshold is a value used to distinguish between "benign" and "anomalous" data. During model testing, by comparing the remaining errors produced by the model on "anomalous" data with the threshold, anomalous data can be identified. The threshold plays a crucial role as it allows the model to effectively identify abnormal data and minimize false positives or false negatives. Determining an appropriate threshold is typically based on the data type and its features, and it might vary across different problem domains.

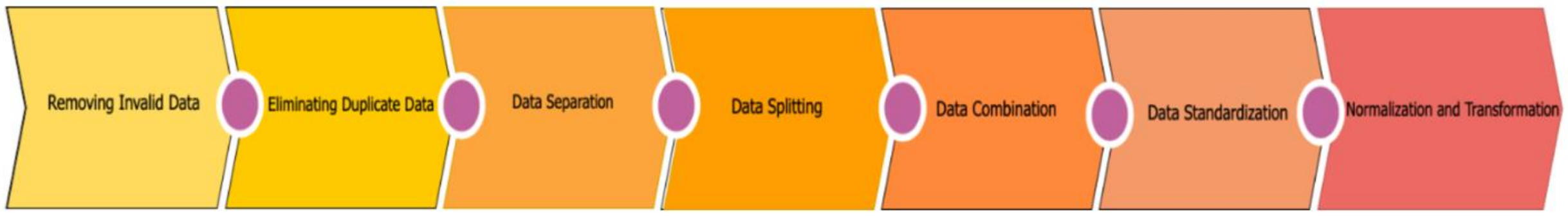

**Figure 6.** Data Preprocessing and cleaning steps

5) Data Combination:

From the "anomalous" data, random sampling is conducted to create a blend of "benign" and "anomalous" data, which will be used for model testing.

6) Data Standardization:

"Benign" data used for model training and threshold calculation are standardized. Standard scaling is applied using the standard scaler for this purpose.

7) Normalization and Data Transformation:

"Benign" data used for threshold calculation are transformed into PyTorch tensors and placed on the utilized device (GPU). Additionally, the combined "benign" and "anomalous" data used for model testing are also converted into PyTorch tensors.

**Table 4.** Each Device type and shape after preprocessing and cleaning

| **Device Number** | **Device Name** | **Device Shape** |
|---|---|---|
| Device1 | Danmini - Doorbell | (1018298,117) |
| Device2 | Ennio - Doorbell | (835876,117) |
| Device3 | Ecobee - Thermostat | (355500,117) |
| Device4 | Philips B120N/10 - Baby Monitor | (1098677,117) |
| Device5 | Provision PT-737E - Security Camera | (828260,117) |
| Device6 | Provision PT-838 - Security Camera | (836891,117) |
| Device7 | Simple Home XCS7-1002-WHT - Security Camera | (375222,117) |
| Device8 | Simple Home XCS7-1003-WHT - Security Camera | (863056,117) |
| Device9 | Samsung SNH 1011 N - Web cam | (850826,117) |

### *6.3.*Experimental Setup

All the experiments conducted in this study were performed according to the specific settings outlined in Table 5.

**Table 5.** The experimental environment

| **Software and Hardware** | **Configuration** |
|---|---|
| CPU | Intel Core i9 12900K (24 virtual cpu) |
| RAM | 60GB |
| GPU | Nvidia 3090 Ti |
| HARD | 80GB OS disk |
| OS | Windows 10, 64bit |
| CUDA TOOLKIT | 11.2 |
| cudnn | 8.1.0 |
| Pytorch | 2.0 |
| Python | 3.9 |

## 6.4. Evaluation Metrics

We used a wide range of evaluation indicators to determine the extent to which our proposed FL framework reduces DDoS attacks in IoT networks. These criteria are well-established in the fields of machine learning and cybersecurity and are used to measure the effectiveness of our strategy. We utilized the following evaluation measures in our experiments:

**Accuracy:**

Accuracy assesses the overall correctness of classification findings by indicating the percentage of correctly predicted examples out of all instances.

**Precision:**

Precision is a metric that quantifies the proportion of accurate positive predictions out of all positive predictions, providing insight into the model's ability to reduce false positives.

**Recall:**

Recall measures the proportion of true positive predictions out of all actual positive instances, reflecting the model's ability to identify true positive cases.

**False Positive Rate (FPR):**

The False Positive Rate calculates the proportion of false positive predictions out of all actual negative instances, indicating the model's ability to correctly classify negative instances.

**Specificity:**

Specificity measures the proportion of true negative predictions out of all actual negative instances, representing the model's ability to correctly identify negative cases.

**Negative Predictive Value (NPV):**

The Negative Predictive Value calculates the proportion of true negative predictions out of all negative predictions, offering insights into the model's accuracy in avoiding false negatives.

**Area Under the Curve (AUC):**

The AUC metric assesses the efficiacy of the classification model across various threshold settings. It provides a measure of the model's ability to distinguish between positive and negative instances.

**F1 Score:**

The F1 Score is the harmonic mean of precision and recall, offering a fair assessment of the model's performance in dealing with false positives and false negatives. The evaluation indicators were selected to offer a thorough assessment of the FL-based framework's effectiveness in reducing DDoS attacks in IoT environments. Examining these measures provides us with useful insights into the advantages and constraints of our suggested method.

Furthermore, these measurements assist in comparing our framework's efficacy with conventional DDoS mitigation strategies and in comprehending the influence of FL on the overall security of IoT networks. Here, we will discuss the outcomes of our experiments and provide a thorough examination of the assessment metrics to demonstrate how the FL-based framework improves IoT network security and reduces DDoS attacks.

### 6.5.Results

We have conducted various experimental combinations to evaluate the performance of the proposed federated learning model. In this research, we compare the proposed federated autoencoder model with two aggregation algorithms, as shown in Fig. 7, FedAvgM and FedAvg. For evaluating the performance of FL models, the condition of uniform client selection is maintained, and parameters are presented in Table 1. All devices undergo training, and the average values of positive prediction accuracy, true positive rate, false positive rate, AUC, F1 score, specificity, NPV, and AUC are calculated across nine devices.

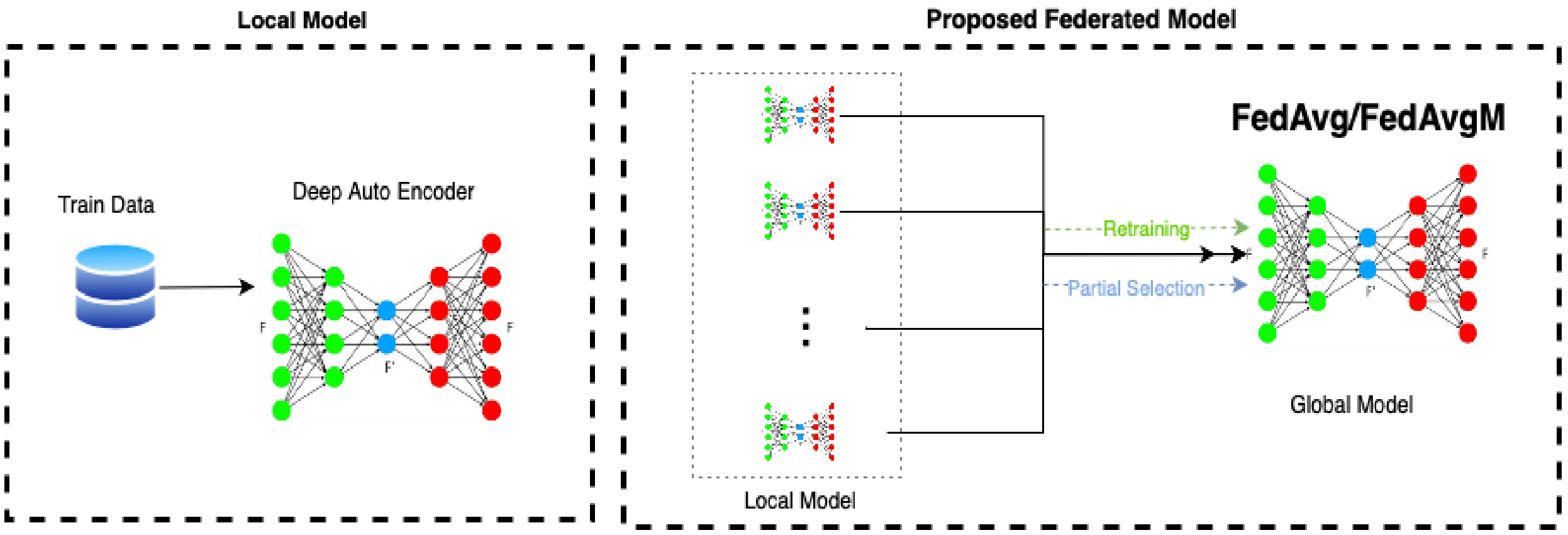

**Figure 7.** Research Scheme

**FedAvg Model**

In general, the results (Fig. 8 and Table 6) demonstrate that the proposed federated learning model, when utilizing the FedAvg aggregation algorithm, exhibits significant proficiency in detecting DDoS attacks. With a remarkably high accuracy, the model succeeds in accurately identifying the majority of data instances and making highly accurate positive predictions.

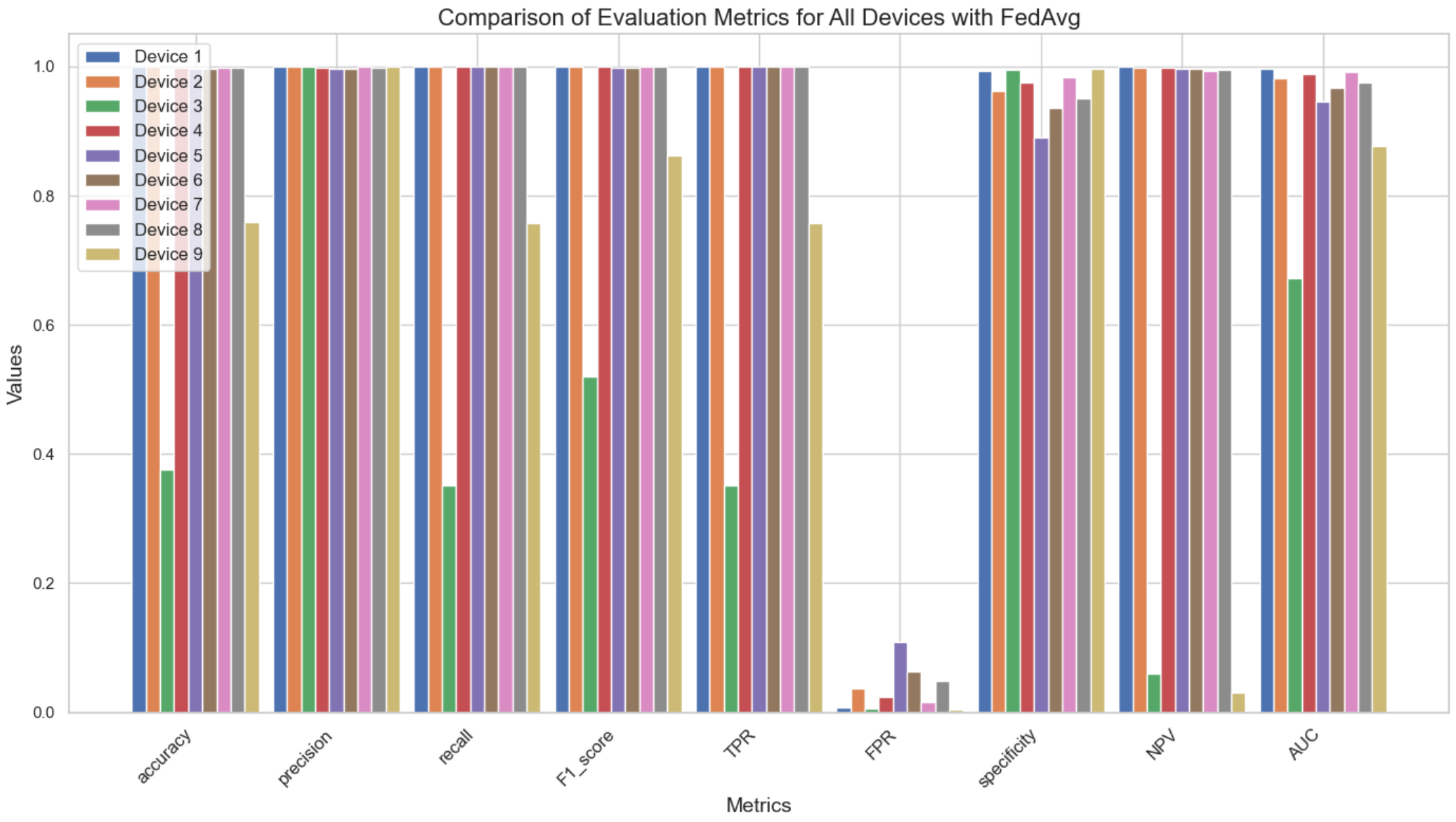

**Figure 8.** Comparison of Evaluation metrics for the proposed model utilizing FedAvg

For all devices, except for Devices 3 and 9, the results reveal that the model achieves an accuracy of over 99.9%, accurately distinguishing between attack and non-attack data. The positive prediction accuracy is also remarkably high, indicative of the model's capability to predict attacks accurately. Moreover, the recall rate is substantially high, suggesting the model's adeptness at detecting and retrieving most attack instances. The F1 score is also close to 1, indicating a balance between precision and recall. The specificity is also notably high, signifying the model's ability to identify a few non-attack samples as attacks. These outcomes emphasize the model's competence in correctly identifying and retrieving both attack and non-attack instances.

**Table 6.** Evaluation Metrics for all 9 devices using the proposed model with FedAvg

| Evaluation Metric / Device# | Accuracy | Precision | Recall | Score F1 | TPR | FPR | Specificity | NPV | AUC |
|---|---|---|---|---|---|---|---|---|---|
| **Device 1** | 0.999 | 0.999 | 0.999 | 0.999 | 0.999 | 0.007 | 0.992 | 0.999 | 0.996 |
| **Device 2** | 0.999 | 0.999 | 0.999 | 0.999 | 0.999 | 0.033 | 0.996 | 0.999 | 0.983 |
| **Device 3** | 0.376 | 0.999 | 0.350 | 0.519 | 0.350 | 0.005 | 0.994 | 0.059 | 0.672 |
| **Device 4** | 0.998 | 0.998 | 0.999 | 0.999 | 0.999 | 0.024 | 0.975 | 0.997 | 0.987 |
| **Device 5** | 0.997 | 0.997 | 0.999 | 0.998 | 0.999 | 0.102 | 0.897 | 0.996 | 0.984 |
| **Device 6** | 0.997 | 0.997 | 0.999 | 0.998 | 0.999 | 0.064 | 0.935 | 0.997 | 0.967 |
| **Device 7** | 0.998 | 0.999 | 0.999 | 0.999 | 0.999 | 0.015 | 0.984 | 0.992 | 0.992 |
| **Device 8** | 0.998 | 0.999 | 0.999 | 0.999 | 0.999 | 0.049 | 0.950 | 0.994 | 0.975 |
| **Device 9** | 0.759 | 0.999 | 0.758 | 0.862 | 0.758 | 0.0033 | 0.996 | 0.031 | 0.877 |

For Devices 3 and 9, results are significantly lower than the other devices. Accuracy and recall are much lower than the average of other devices. This reduction in model performance can be attributed to two factors:

1. **Insufficient Training Data:** Devices 3 and 9 seem to have been allocated fewer training samples, especially those related to all types of attacks. This shortage could lead to a decreased power of the model to detect attacks.

2. **Data Representation Imbalance:** In Devices 3 and 9, the training data might not be well representative of all types of attacks. The model's detection capability might suffer due to the lack of adequate representation of all attack types.

In conclusion, the evaluation results of the federated learning model for DDoS attack detection using the FedAvg aggregation algorithm demonstrate that the model achieves a highly accurate detection of attack and non-attack instances, with a precision rate of approximately 99.9%. The positive prediction accuracy is also extremely high, while the recall rate is also around 99.9%, indicating the model's ability to detect and retrieve the majority of attack instances. The F1 score, true positive rate, and false positive rate all reflect the model's ability to distinguish between attack and non-attack events. The high specificity further demonstrates the model's ability to classify a few non-attack samples as attacks.

**FedAvgM Model:**

In relation to the FedAvgM algorithm, all the considerations mentioned for FedAvg hold true, with the distinction that a comparison between Table 6 and Table 7 reveals the considerably superior performance of FedAvgM across all devices compared to the FedAvg algorithm. The FedAvgM algorithm requires notably less training time compared to FedAvg. This reduction signifies an improved training process efficiency in FedAvgM. The decreased training time implies optimized computational resource utilization and cost reduction.

**Table 7.** Evaluation Metrics for all 9 devices using the proposed model with FedAvgM

| Evaluation Metric / Device# | Accuracy | Precision | Recall | Score F1 | TPR | FPR | Specificity | NPV | AUC |
|---|---|---|---|---|---|---|---|---|---|
| **Device 1** | 0.999 | 0.999 | 0.999 | 0.999 | 0.999 | 0.007 | 0.992 | 0.999 | 0.996 |
| **Device 2** | 0.999 | 0.999 | 1.0 | 0.999 | 1.0 | 0.041 | 0.958 | 1.0 | 0.979 |
| **Device 3** | 0.376 | 0.999 | 0.350 | 0.519 | 0.350 | 0.004 | 0.995 | 0.0059 | 0.673 |
| **Device 4** | 0.998 | 0.998 | 0.999 | 0.999 | 0.999 | 0.022 | 0.977 | 0.997 | 0.988 |
| **Device 5** | 0.999 | 0.999 | 0.999 | 0.999 | 0.999 | 0.026 | 0.973 | 0.997 | 0.986 |
| **Device 6** | 0.998 | 0.998 | 0.999 | 0.999 | 0.999 | 0.028 | 0.971 | 0.997 | 0.985 |
| **Device 7** | 0.998 | 0.999 | 0.999 | 0.999 | 0.999 | 0.015 | 0.984 | 0.992 | 0.992 |
| **Device 8** | 0.999 | 0.999 | 0.999 | 0.999 | 0.999 | 0.047 | 0.952 | 0.995 | 0.976 |
| **Device 9** | 0.759 | 0.999 | 0.758 | 0.862 | 0.758 | 0.003 | 0.996 | 0.031 | 0.877 |

The reduction in training time with FedAvgM contributes to a lower probability of model overfitting. FedAvgM excels in harnessing training data to cultivate a model endowed with enhanced and more expansive learning capabilities. This enhancement leads to improved attack detection. The decreased training time in FedAvgM also translates to reduced computational costs. This aspect can elevate the quality and efficiency of the training stages, providing results similar to or better than FedAvg while utilizing fewer resources. The subsequent section elaborates on the detailed comparison between the two models.

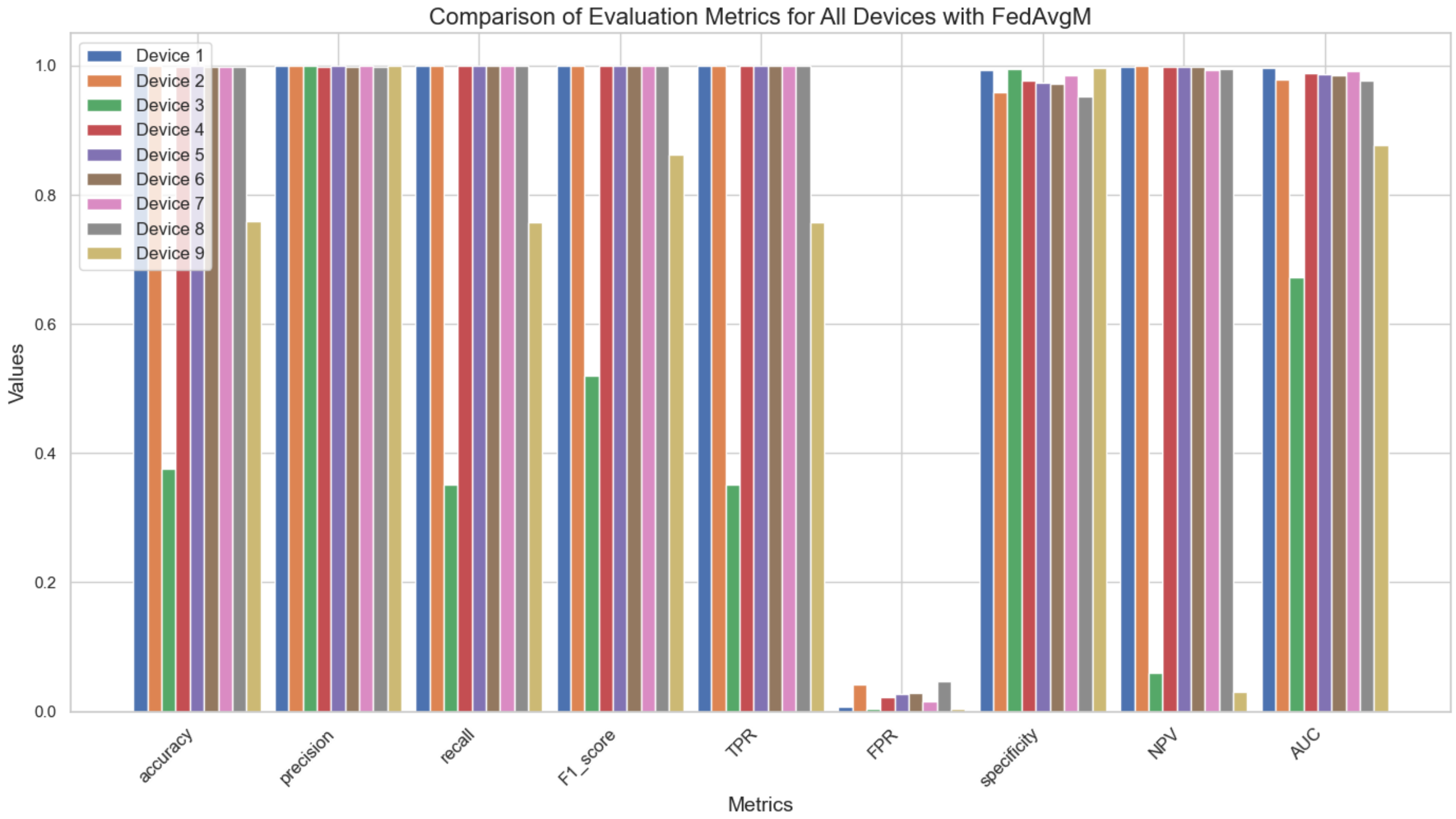

**Figure 9.** Comparison of Evaluation metrics for the proposed model utilizing the FedAvgM

**Comparison between FedAvg and FedAvgM:**

Figures 9 and 10 illustrate the evaluation results of the proposed federated model using both FedAvg and FedAvgM algorithms. Both algorithms provide excellent performance in a variety of evaluation criteria, including accuracy, positive prediction accuracy, recall, true positive rate, false positive rate, specificity, negative prediction accuracy, and ROC-AUC. However, as highlighted in Table 8, subtle differences exist between these two algorithms, indicating which one is more suitable for the nature of our research. From Table 8, it's evident that the mean false positive rate for FedAvgM is lower than that of FedAvg, demonstrating a better detection capability of FedAvgM for correct attacks. Moreover, the mean F1 score for FedAvgM is higher, indicating that FedAvgM achieves a superior balance between positive prediction accuracy and coverage. Taking these parameters and other metrics presented in Figure 10 into consideration, it can be concluded that the performance of FedAvgM is superior, and its adoption contributes to the enhancement of FL model performance.

**Table 8.** Comparison of the Proposed model using FedAvg and FedAvgM

| Model Type | Precision Avg | TPR Avg | FPR Avg | F1-score Avg | AUC Avg | Time (sec) |
|---|---|---|---|---|---|---|
| FedAvg | 99.89% | 90.09% | 3.40% | 93.07% | 93.34% | 58.43 |
| FedAvgM | 99.93% | 90.09% | 2.19% | 93.10% | 93.95% | 49.01 |

## 7. Conclusion

We performed a thorough assessment of the suggested methodology for detecting DDoS attacks in IoT networks. The main goal of this study was to thoroughly analyze the performance of the suggested model while exposed to different forms of DDoS attacks. The model's performance was evaluated by analyzing the findings from nine different devices in the federated learning environment. The evaluation criteria used in this study include accuracy, positive prediction accuracy, recall, F1-score, true positive rate, false positive rate, specificity, NPV, and AUC.

Given the significant data distribution imbalance between normal and abnormal data in this study, where most devices exhibited substantial volumes of abnormal traffic data and scarcity of normal data, particular devices like D1, D2, and D8 demonstrated notable improvements in the proposed model's performance with exceptionally high accuracy. This highlights the model's potential to detect attacks and non-attacks across these devices. Additionally, devices D4, D5, D6, and D7 also exhibited substantial effectiveness in detecting attacks. These findings suggest that the model consistently operated across various devices in detecting DDoS attacks and showcased the ability to detect such attacks uniformly. However, devices D3 and D9 exhibited lower accuracy, possibly due to an insufficient amount of data encompassing all types of DDoS attacks for these devices. This observation accentuates the reliance of attack detection accuracy on the quality and type of training data available for each device.

Since our model operates as a binary classifier, our most important metrics among all the utilized ones are the AUC and F1-score. Therefore, all adjustments and hyperparameter tuning primarily focused on improving these two metrics. This research allowed us to evaluate the capability and performance of this approach against security challenges posed by DDoS attacks and emphasize the importance of optimal execution and regular update cycles of federated learning. The high accuracy and composite metrics of the AUC and F1-score demonstrate the proposed model's proficiency in detecting attacks with precision and efficiency. The final evaluation confirms the proposed framework as an effective security solution against DDoS attacks in IoT networks. It also implies that this approach facilitates collaboration and data sharing among devices without the need to transmit sensitive data to a central entity, thus enhancing the security and federated performance of IoT networks.

## 8. Discussion & Future Work

One of the main limitations we encountered was the heterogeneity of data in the N-BaIoT dataset. Traditional federated learning assumes that training data are shared across different devices in a common feature space, yet this assumption might not hold in real-world scenarios. Our Dataset encompasses diverse devices, including security cameras, doorbells, and thermostats. Transfer federated learning enables models trained on a large dataset in one domain to be utilized in other related domains.

Moreover, in this paper, we employed two aggregation methods for federated learning, namely FedAvg and FedAvgM. However, FedAvg is not a comprehensive solution and makes simplifying assumptions. For instance, it assumes that all sampled devices complete a certain number of local stochastic gradient descent iterations (e.g., e epochs), but this assumption may not hold for all devices. Slow-moving devices could hinder the convergence speed. Even though FedAvg prunes certain clients from the client list, if 90% of the devices are slow-moving, the model's performance deteriorates. Furthermore, its performance is not guaranteed for all types of heterogeneous data. As a result, more sophisticated aggregation algorithms such as The Federated Matching Algorithm (FedMA) or FedProx can be utilized for further research[47], [48].

In conclusion, there are multiple avenues for improving federated learning algorithms to accommodate heterogeneous data and enhance performance. Methods like Transfer federated Learning, FedMA, and FedProx are promising directions for future research to address the limitations and challenges posed by real-world data heterogeneity.